# Implications of Neutrino Balls as the Source of Gamma-Ray Bursts

*D. Syer*

Canadian Institute for Theoretical Astrophysics, MacLennan Labs, 60 St. George Street, Toronto M5S 1A7, Ontario.

# Summary

Holdom and Malaney (1994) have suggested a mechanism for gamma-ray bursts which requires that stars be captured by a neutrino ball. Neutrino balls would be, for the most part, denser than main sequence stars, but their density would decrease as their mass increased. We show that small neutrino balls would subject stars to tidal forces sufficient to disrupt them. We thus argue that if neutrino balls existed at the centres of galaxies, only the largest would be able to act as a source of gamma-ray bursts. Such neutrino balls would have a mass of order $10^7 M_\odot$. Tidal capture of stars by a neutrino ball would not be important, but dynamical friction against the neutrinos or star-disc interactions could both be important capture mechanisms. We find that a gamma-ray burst would occur in a galaxy containing such a neutrino ball roughly every $10^2$ y, and the fraction of all galaxies contributing to the gamma-ray burst flux would be $\sim 10^{-4}$, assuming that this was the mechanism of all gamma-ray bursts. These numbers have implications for neutrino ball models of active galaxies, assuming that all gamma-ray bursts and all AGN come from neutrino balls. Either a small fraction $\sim 10^{-2}$ of the lifetime of such an object could be spent as an AGN, or that the probability of a neutrino ball becoming an AGN would be $10^{-2}$. It is not possible to rule out the possibility that neutrino balls might exist at the centres of galaxies through direct ground-based observation of stellar kinematics.

# 1 Introduction

Holdom and Malaney (1994) describe a mechanism for the production of gamma-ray bursts which involves a supernova going off inside a neutrino ball. Neutrino balls are 'cosmic balloons' (Holdom 1993), consisting of right handed neutrinos confined by a spherical domain wall. Outside the domain wall right-handed neutrinos are supposed to exist and have masses of order 1 TeV, and left-handed (normal) neutrinos have very small masses. Inside the domain wall, chirality is exchanged and right-handed neutrinos have very small masses. This results in a surface tension, $\gamma$, in the domain wall which confines a gas of right-handed neutrinos inside a neutrino ball. The cooling of a star would be greatly enhanced inside a neutrino ball, which would ultimately be responsible for a supernova. The neutrino blast wave from a supernova would interact with anti-neutrinos in the neutrino ball, creating relativistic $e^+e^-$ pairs. The blast wave would lose a fraction of order

$$\alpha \sim 10^{-3}(\gamma/\,\mathrm{TeV}^3) \qquad (1)$$

of its energy in passing through the neutrino ball, most of which would come out as gamma rays.

There are few *a priori* constraints on the properties of neutrino balls. There is a lower mass bound below which a neutrino ball would evaporate, and an upper bound above which it would become self gravitating and collapse to form a black hole. At the upper end of the mass range, a neutrino ball would be almost self-gravitating, so the escape speed from its surface, $v$, would be of order the speed of light, $c$. In terms of its surface tension, $\gamma$, the mass of a neutrino ball satisfies

$$10^4 \gamma^3 M_\odot \lesssim M_\mathrm{b} < 2 \times 10^7/\gamma M_\odot \qquad (2)$$

for $\gamma$ in TeV$^3$ (Holdom 1987, Holdom 1993). The balance between constant surface tension and pressure demands that the density of a neutrino ball is roughly inversely proportional to its radius.

The questions which remain to be settled include: how a star gets to be inside a neutrino ball; and what, if anything, might constrain the distribution of neutrino balls, assuming that they are responsible for all gamma-ray bursts. The first question is the main subject of this paper and is dealt with in the next section. The second question is the subject of Section 3, in which we also estimate the gamma-ray burst rate from a single neutrino ball.

## 2 Catch a Falling Star...

We will now show that, for $\gamma = 1\,\mathrm{TeV}$, all of the neutrino balls allowed by (2) would actually tidally strip a star, rather than capturing it whole. The distance from the centre of a neutrino ball at which a star would be tidally stripped is $R_\mathrm{T}$ where

$$R_\mathrm{T} \simeq 1.3 r_* \left(\frac{M_\mathrm{b}}{m_*}\right)^{\frac{1}{3}}, \qquad (1)$$

where $r_*$ is the radius of the star and $m_*$ is its mass (Carter and Luminet 1985). The tidal force would be maximum at the surface of the neutrino ball, so it is sufficient to examine the ratio $\eta = R_\mathrm{T}/R_\mathrm{b}$ to determine whether a star can be captured whole. Setting $M_\mathrm{b}$ equal to the maximum mass allowed by (2), in which case $R_\mathrm{b} \simeq 3.85 G M_\mathrm{b}/c^2$ (Holdom 1993), we find that

$$\eta = \frac{R_\mathrm{T}}{R_\mathrm{b}} \simeq 1.8 \frac{r_*}{r_\odot} \frac{M_\odot}{m_*} \left(\frac{\gamma}{\mathrm{TeV}^3}\right)^{-\frac{2}{3}}. \qquad (2)$$

Thus the tidal limit would be (marginally) outside the most massive neutrino balls for $\gamma = 1\,\mathrm{TeV}^3$. This is potentially the most severe problem for the neutrino ball model of gamma-ray bursts. There is enough uncertainty in the theory of neutrino balls, however, that it might nevertheless be productive to examine the consequences of the model.

For instance, decreasing $\gamma$ would make $R_\mathrm{T} < R_\mathrm{b}$, and hence allow stars to be captured whole, but it would also decrease the gamma-ray burst energy (equation 1). Two other physical mechanisms act in the right direction to decrease the risk of tidal disruption: firstly, the core of a star can still produce a supernova, so provided only the outer parts are tidally removed, gamma-ray bursts of sufficient strength might still be produced. Only ten or twenty percent at most of a star could be stripped before its average density actually decreased (Hjellming and Webbink 1987). We might hope for an increase in the density of a star of a factor of two, gaining a factor of $2^{1/3}$ in the tidal radius. Secondly, the tidal force on a star would be maximum at the surface of a neutrino ball (compare the tidal force due to a point mass, which increases indefinitely to smaller distances). A star on an orbit that intersected a neutrino ball would only be exposed to strong tidal forces for a short time (less than a sound crossing time in the star) and might therefore survive the encounter. We will assume that the uncertainties (particularly in equation 1) can conspire to make gamma-ray bursts in neutrino balls at the top end of (2), with masses $\sim 10^7 M_\odot$, and that those which are less massive would tidally disrupt even the core of a star.

A star on a radial orbit would not be inside a neutrino ball for a time significantly in excess of its sound-crossing time. It would therefore not be able to reach a new hydrostatic equilibrium after the changes induced by the enhanced cooling that it would suffer in one pass through a neutrino ball. The cooling, however rapid, could therefore not be the cause of a supernova in this period. Thus for a star to give rise to a supernova inside a neutrino ball it must first be gravitationally bound to it. For this reason we next consider the mechanisms by which a star may be captured by a neutrino ball, that is, by which it may become gravitationally bound to the neutrino ball. The same mechanisms would be responsible for the star eventually becoming embedded in a neutrino ball, as required for the production of gamma-ray bursts. We will then return to the possibility of supernovae inside a smaller neutrino ball. But first, we give a description of the stellar-dynamic state of the centre of a galaxy, including the influence of a neutrino ball.

We may readily envision a neutrino ball sitting in the centre of a galaxy. Tremaine, Ostriker and Spitzer (1975) calculated that globular clusters would be efficiently dragged into the centre of a galaxy by dynamical friction. A neutrino ball, more massive than a globular cluster, would be dragged into the centre by dynamical friction on the stars in a an even shorter time. Once in the centre of a galaxy a neutrino ball would attract stars gravitationally. It would be surrounded by a dense cusp of stars inwards from a radius $R_\mathrm{c}$ given by

$$R_\mathrm{c} \sim \frac{GM_\mathrm{b}}{\sigma^2} \qquad (3)$$

where $\sigma$ is the velocity dispersion in the ambient star cluster. Provided $R_\mathrm{c}$ is greater than any homogeneous core that the cluster had, then $\sigma$ would be given

roughly by

$$\sigma^2 \sim \frac{GM(R)}{R} \qquad (4)$$

where $M(R)$ is the mass of stars within a radius $R$ of the centre. Setting $R = R_c$ in (4) we see that

$$M_c \sim M_b, \qquad (5)$$

or in other words, the neutrino ball 'grabs' of order its own mass from the star cluster. Outside the cusp the density of stars decreases.

The stars inside $R_c$ can be bound or unbound to the neutrino ball. If the relaxation time in the cusp were short then the majority of stars in the cusp would be bound, and the density would follow an $r^{-7/4}$ profile. In this case, stars would be subjected to disruptive collisions with each other before they reached an orbit at the surface of the neutrino ball. (This follows from the fact that the orbital velocity of the stars would be greater than their internal sound speed. It is independent of the mass of the neutrino ball insofar as all neutrino balls have an escape velocity satisfying this condition.) All the stars in the cusp around a neutrino ball would be unbound, and the density would follow roughly a relatively weakly varying profile, as determined by Liouville's Theorem. If the undisturbed star cluster had a constant density inside $R_c$, the resulting density is proportional to $r^{-1/2}$ (Duncan and Shapiro 1983).

We will be interested in the scattering of stars by mutual encounters because they can be scattered into orbits which come close to the neutrino ball. Each star is scattered so that its angular momentum with respect to the centre of the cluster changes in a relaxation time, given by

$$\tau \sim \frac{1}{4\pi\Lambda} \frac{\sigma^3}{G^2 \rho m_*} \qquad (6)$$

where $\rho$ is the local density of stars and $\Lambda$ is the Coulomb logarithm (Binney and Tremaine 1987). Using equations (5) and (3) and writing $\rho \sim M(R)/R^3$, the relaxation time of stars at the edge of the cusp would thus be given by

$$\tau_c \sim \frac{1}{4\pi\Lambda} \frac{G^2 M_b^2}{\sigma^3}. \qquad (7)$$

For the present purposes $\Lambda$ will be taken to be of order unity because nowhere in the system would the density be homogeneous over a large range of radius, and hence we are interested in relaxation which is in some sense local. For $M_b \sim 10^7 M_\odot$ and $\sigma \sim 200\,\mathrm{km\,s^{-1}}$ we find that $\tau_c \gtrsim 10^{10}$ y. This is weak *a posteriori* justification for the assumption that $R_c$ lies outside the homogeneous core of the unperturbed cluster: had our estimate for $\tau_c$ been substantially less then the cluster would have been strongly evolving on a cosmological timescale. Data from the Hubble Space Telescope may indicate that homogeneous cores in galaxies are rare in any case (Crane *et al.* 1993).

To be captured into the centre of the cluster, a star must be subjected to a drag force. A neutrino ball might provide such a force in three different ways: energy transfer by tidal forces; dynamical friction on the star by the neutrino gas; or interaction between the star and a gas disc around the neutrino ball. We shall conclude that the second and third mechanisms would be most important for capture, and dynamical friction would be important for dragging a star into the centre of a neutrino ball. Holdom and Malaney (1994) argue that a supernova at the centre of a neutrino ball would give rise to a gamma-ray burst with a characteristic echo signature.

Tidal capture of a star by a neutrino ball might be the dominant mechanism, given the arguments above that tidal forces are appreciable (equation 2). A star might be captured in only a few orbits, but in order to be tightly bound or for its orbit to be inside the neutrino ball, it would have to absorb more than its own self-binding energy. It would not have time to radiate this energy since the orbital timescale would be much shorter than the thermal timescale of the star, and thus it would be disrupted by the cumulative effect of many tidal encounters (Syer, Clarke and Rees 1991).

Dynamical friction against the neutrino fluid inside a neutrino ball was briefly discussed by Holdom and Malaney (1994) as a way of winding the star into the centre of a neutrino ball, once it was captured. The Coulomb logarithm would of order unity because a neutrino ball would be not many orders of magnitude larger than a star. A star originally marginally unbound or weakly bound to the neutrino ball would move at velocity $v \sim (GM_{\rm b}/R_{\rm b})^{1/2}$ through the neutrino ball. For a self-gravitating neutrino ball of mass $\sim 10^7 M_\odot$, $v \sim c$. Writing $\rho_{\rm b}$ for the density of the neutrino ball, the star feels an acceleration

$$\dot{v} \sim \frac{m_*}{M_{\rm b}} \left(\frac{v}{c}\right)^3 G(\rho_{\rm b})^{1/2} v$$

(Binney and Tremaine 1987) and loses specific energy $\Delta E \sim R_{\rm b} \dot{v}$ in one orbit. The number of such orbits a star would have to make in order for it to become bound to the neutrino ball is

$$n_{\rm bind} \sim \frac{\sigma^2}{\Delta E} \sim \frac{M_{\rm b}}{m_*} \left(\frac{c}{v}\right)^3 \left(\frac{\sigma}{v}\right)^2. \tag{8}$$

Since $v \propto M_{\rm b}^{1/4}$, and $v \sim c$ for $M_{\rm b} \sim 10^7 M_\odot$, we have

$$n_{\rm bind} \sim 10^2 \left(\frac{10^7 M_\odot}{M_{\rm b}}\right)^{1/4} \left(\frac{\sigma}{200\,{\rm km\,s^{-1}}}\right)^2. \tag{9}$$

Once it was bound, the star would quickly become embedded in the neutrino ball, shrinking its orbit by a factor of order unity every $n_{\rm bind}$ orbits, with the orbital timescale ever decreasing.

We must check to make sure that in the process of becoming bound to the neutrino ball, the star would not be scattered away from its radial orbit. Problems of capture from a star cluster by a process with $n_{\rm bind} \gtrsim 1$ are equivalent to the star-disc capture mechanism of Syer, Clarke and Rees (1991). In such a situation the usual concept of a 'loss-cone' (Frank and Rees 1976), from which a star will be captured in one crossing time has to be modified slightly. Suppose the star would be scattered out of its orbit intersecting a neutrino ball in $n_{\rm diff}$ passes, then to be captured a star must be in an orbit for which $n_{\text{diff}} \gtrsim n_{\text{bind}}$. Stars satisfying this condition are said to inhabit a 'meta-loss cone.' The existence of the meta-loss-cone at the edge of the cusp $R_c$ would ensure that there be capture by dynamical friction. The capture rate is determined by the location $R_{\text{crit}}$ at which $n_{\text{diff}} \sim n_{\text{bind}}$, and is of order $M(R_{\text{crit}})/\tau(R_{\text{crit}})$.

We assume that the presence of a central point mass would be sufficient to quickly smooth out the star cluster and make it spherical within $R_c$ (see Gerhard and Binney 1987). The dominant source of scattering would then be interactions with other stars, as opposed to large-scale tidal effects. In one relaxation time, the angular momentum of the star, $l \sim R_b v$, would diffuse by of order $L \sim R_c \sigma$, the angular momentum of a circular orbit with the same energy. Thus, the time for $l$ to change by of order unity is

$$t_l \sim \left(\frac{l}{L}\right)^2 \tau \sim \frac{M_b^2}{M_c^2}\left(\frac{\sigma}{v}\right)^2 \tau \qquad (10).$$

The number of orbits before the angular momentum of the star was changed by of order unity would be

$$n_{\text{diff}} \sim \frac{M_b^2}{M_c m_*}\left(\frac{\sigma}{v}\right)^2. \qquad (11)$$

Requiring that $n_{\text{diff}} \gtrsim n_{\text{bind}}$ for the existence of a meta-loss cone, we find that

$$\frac{M_b}{M_c}\left(\frac{v}{c}\right)^3 \gtrsim 1 \qquad (12)$$

which, with $M_b \sim M_c$ would be marginally true for a self-gravitating neutrino ball. Thus $R_{\text{crit}} \sim R_c$ and the capture rate would be dominated by scattering into the meta-loss-cone at $R_c$. Note that (12) is independent of $\sigma$—this is a general feature of meta-loss cone problems.

The tidal energy transfer in one pass is of order $\eta^6$ times the self-binding energy of a star, and is thus very sensitive to the uncertainties in the value of $\eta$. If $n_{\text{bind}} \gg \eta^{-6}$, stars which otherwise would be subject to capture by dynamical friction would be tidally disrupted before they became tightly bound to a neutrino ball. But given the sensitivity of the tidal energy transfer to $\eta$, that $\eta < 1$ in order that a star survive even a single pass, and that $n_{\text{bind}}$ is not $\gg 1$, we conclude that dynamical friction might be an important mechanism binding stars to a neutrino ball.

Interactions between a star and a disc around a super-massive black hole were discussed by Syer, Clarke and Rees (1991). The problem of whether or not stars are captured by a disc is a meta-loss-cone problem, similar to that of capture by dynamical friction. For drag against a disc the equivalent of (12) would be a minimum surface density in the disc:

$$\Sigma_{\text{crit}} \sim \frac{m_*}{M_b}\frac{m_*}{r_*^2} \qquad (13)$$

for $M_c \sim M_b$. A star could also be deposited in a bound orbit by the tidal disruption of a binary star. The star need not be captured close to the neutrino ball, but there are some additional conditions that ensure the star would be swept down to the surface of the neutrino ball by the disc: a minimum thickness and a maximum viscosity in the disc, without which the star cannot form a gap in the accretion disc. An additional condition would be that the mass in the disc in the region of the neutrino ball is greater than a solar mass.

The self-gravitating neutrino balls that we are interested in would lie inside their last stable circular orbit, so the disc would have non-circular orbits in a region inside the last stable orbit and outside the radius at which a circular disc forms again inside the neutrino ball. There would be very little shear in the disc inside the neutrino ball because the density of neutrinos would be almost constant, so the disc would be of a different character, but dynamical friction would drag the star into the centre, independent of its interaction with the disc.

For a smaller, denser neutrino ball, a star may not need to be captured in order to produce a supernova inside the ball. The possibility of detonating a supernova in a star by a close pass inside the tidal radius of a super-massive black hole has been discussed by Carter (1992) and Carter and Luminet (1985). The mechanism involves squeezing the star tidally in a direction perpendicular to the orbit, and it requires a super-massive object in order that the timescale of the squeezing would be fast enough to set off a runaway nuclear fusion reaction in the core of the star. In contrast to the black hole case, a star's orbit can take it inside the neutrino ball, where the squeezing continues. However, the shortest timescale for the squeezing would be the crossing time of the neutrino ball, and this time must be substantially less than the sound crossing time in a star to detonate a supernova. Unfortunately, even the smallest neutrino balls allowed by (2) would only have a density 100 times that of a star, so a star at the surface of such a neutrino ball would only have penetrated its tidal radius by a factor of a few. In such a case Luminet and Pichon (1989) calculated that only $10^{47}$ erg would be generated in enhanced nuclear burning in the star. Compression would not drive a strong enough shock into the star to detonate a supernova. Thus insufficient neutrinos would be emitted to produce a gamma ray burst.

## 3 The Gamma-Ray Burst Rate

When stars are removed from a cluster by some process at the centre, low angular momentum stars are depleted in a 'loss-cone' (Frank and Rees 1976). If we assume that the loss-cone radius was such that it was empty out to a radius greater than $R_c$, then any mechanism which deposited low-angular-momentum stars in the neutrino ball would result in a gamma-ray burst rate of order $M_c/\tau_c$, or several per crossing time, because stars would be scattered into the loss-cone on a timescale $\tau_c$ at the edge of the cusp. (The contribution from larger radii is smaller because as the density decreases, the crossing time increases.) The classical condition that the loss-cone be empty at $R_c$ can be written as $n_{\text{diff}} > 1$. The condition that the meta-loss-cone be empty at $R_c$ can be written as $n_{\text{diff}} > n_{\text{bind}}$, which would be satisfied provided (13) or (12) was. In this case, the capture rate would be essentially independent of the capture mechanism. Clusters of main sequence stars must have $\sigma \lesssim 1000 \, \text{km s}^{-1}$, but a typical value of $\sigma$ might be $200 \, \text{km s}^{-1}$. Thus the gamma-ray burst repetition rate, per galaxy, would be

$$\dot{N}_{\text{burst}} \sim \frac{M_c}{\tau_c} \sim 10^{-2} \left( \frac{200 \, \text{km s}^{-1}}{\sigma} \right)^3 \frac{M_b}{10^7 M_\odot} \, \text{y}^{-1}, \tag{1}$$

given the assumptions above. This is not yet in conflict with observations of gamma-ray bursts, i.e the repeat rate from a single direction in the sky is not known to be significantly greater than $10^{-2} \, \text{y}^{-1}$.

Given that the mass of a neutrino ball that gives rise to gamma-ray bursts must be at the top of the range (2), and assuming that they are the cosmological source of gamma-ray bursts, we can find some constraints on the population of galaxies which contain such objects.

Mao and Paczynski (1992) give the frequency of gamma-ray bursts as $\sim 10^{-6} \, \text{y}^{-1}$ per galaxy. Thus the fraction of galaxies containing neutrino balls of mass $M_b$, using (1), would be

$$f \sim 10^{-6} \dot{N}_{\text{burst}} \sim 10^{-4} \left( \frac{200 \, \text{km s}^{-1}}{\sigma} \right)^3 \frac{M_b}{10^7 M_\odot}. \tag{2}$$

Assuming $1000 \, \text{km s}^{-1} > \sigma > 50 \, \text{km s}^{-1}$, we have $10^{-6} < f < 10^{-2}$.

Dolgov and Martin (1990) have argued that ordinary matter of very high density inside a neutrino ball could produce an Eddington luminosity from the reaction $\nu \bar{\nu} e^- \to e^- \gamma$. They were not aware that the upper limit to the mass of a neutrino ball was as low as $10^7 M_\odot$, and we find that this has fatal consequences for their argument. Their results imply that the lifetime of a neutrino ball would be

$$\tau_\gamma \sim 10^{11} \left( \frac{50 \text{keV}}{\mu} \right)^8 \frac{\rho_\nu}{\rho_e} \, \text{y}, \tag{3}$$

where $\mu$ is the Fermi energy of the neutrinos, and $\rho_\nu$ and $\rho_e$ are the densities of neutrinos and electrons, respectively. The results of Holdom (1993) imply that

$$M_b \sim 10^7 M_\odot / \gamma \left( \frac{50 \text{keV}}{\mu} \right)^8, \tag{4}$$

which is $10^2$ times smaller than the value quoted by Holdom (1987), and that used by Dolgov and Martin. Combining (4) with (3) we find that

$$\tau_\gamma \sim 10^{11} / \gamma \frac{M_b}{10^7 M_\odot} \frac{\rho_\nu}{\rho_e} \, \text{y}. \tag{5}$$

Comparing this with the Eddington timescale, which is $\sim 10^9$ y for efficient conversion of mass to energy, we see that, to emit an Eddington luminosity, a neutrino ball must have inside it a higher average mass density in electrons than neutrinos. Including the fact that the proton density would be $m_p/m_e$ times higher this would lead to gravitational instability of the baryonic matter inside a neutrino ball, and the formation of a baryonic black hole. This would be fatal for the neutrino ball, which, if it were self-gravitating, would be swallowed by the black hole in a light-crossing time. Dolgov and Martin point out that the reaction $\nu\bar{\nu}e^- \to e^-\gamma$ might take place only at the surface of a neutrino ball, but this would only increase $\tau_\gamma$ and require a higher electron density for an Eddington luminosity.

We might also expect that a large neutrino ball might have an accretion disc around it and act, from the point of view of quasar activity, in the same way as a black hole of the same mass. The specific frequency of $10^7 M_\odot$ neutrino balls deduced above is greater than the specific frequency of active galaxies, which is $\sim 10^{-6}$. Thus if the same galaxies which are or were AGN also correspond to the gamma-ray burst population, the AGN must last for a fraction $10^{-6}/f$ of the life of such an object. For $\sigma \simeq 200\,\mathrm{km\,s^{-1}}$ the fraction would be $10^{-2}$.

The observed positions of gamma-ray bursts are not associated with those of AGN. Assuming the standard model for active galaxies, a large neutrino ball with an accretion disc accreting at its Eddington limit would pass for an AGN. There is an apparent contradiction in these two facts, since we might expect a neutrino ball in an AGN phase to exhibit gamma-ray bursts by the star-disc capture mechanism. To explain this, we observe that such a neutrino ball would accrete of order its own mass in baryons on the Eddington timescale $\sim 10^8$ y, which is short compared to cosmological timescales. The rapid growth of the mass inside a neutrino ball would expose a star at its surface to even higher tidal forces, the effect of which would be to switch off gamma-ray burst activity. A neutrino ball accreting at the Eddington rate would gorge itself on baryons and a black hole would form inside it, eventually to devour the neutrino ball itself. Thus gamma-ray bursts must come from neutrino balls which have never been able to accrete fast enough to flare up into AGN. In this case, we would conclude that the probability of a neutrino ball becoming an AGN is $10^{-2}$ for $\sigma \simeq 200\,\mathrm{km\,s^{-1}}$, assuming all AGN were at one time neutrino balls.

A point mass this large at the centre of our own galaxy would have been observed through its effects on the stellar cluster around it. But in more distant galaxies it would be harder to detect directly, so the actual fraction of galaxies with directly observable large neutrino balls would be much less than $f$. A cluster would be influenced out to a radius of $R_c \sim M_b/\sigma^2$, which would subtend an angle of $\delta$ arc second at a distance

$$D \sim \frac{0.1''}{\delta} \frac{M_b}{10^7 M_\odot} \left(\frac{200\,\mathrm{km\,s^{-1}}}{\sigma}\right)^2 2\,\mathrm{Mpc}. \tag{6}$$

The number of galaxies observable would be $N_{\mathrm{obs}} \sim f\rho_{\mathrm{gal}}D^3$, where $\rho_{\mathrm{gal}}$ is the co-moving density of galaxies. Estafthiou $et\,al$ (1988) give $\rho_{\mathrm{gal}} \sim 10^{-2}\,\mathrm{Mpc}^{-3}$, implying

$$N_{\mathrm{obs}} \sim 10^{-5} \left(\frac{0.1''}{\delta}\right)^3 \left(\frac{M_b}{10^7 M_\odot}\right)^2 \left(\frac{200\,\mathrm{km\,s^{-1}}}{\sigma}\right)^3. \tag{7}$$

The local density of galaxies is closer to $1\,\mathrm{Mpc}^{-3}$, but even then we would not expect to observe a neutrino ball using a telescope with $\sim 0.1$ arc second resolution.

At least two galaxies—M31 (Kormendy 1988, Dressler and Richstone 1988), and M32 (Dressler and Richstone 1988)—have been suggested as having central point masses of $10^7 M_\odot$ or larger as a result of direct ground-based observations. Data from the Hubble Space Telescope may indicate that possibly the majority of galaxies have central density cusps which are not inconsistent with those associated with large point masses. Thus it would be very difficult to rule out the possibility of enough galaxies having large neutrino balls to give rise to all observed gamma-ray bursts, possibly even as a very small subset of galaxies with central masses in the $10^7 M_\odot$ range. From a stellar-kinematic point of view, a neutrino ball at the centre of a galaxy would be indistinguishable from a black hole of the same mass.

# References

Carter, B. 1992, *Astrophys. J.*, **391**, L67.
Binney, J., and Tremaine, S. 1987, *Galactic Dynamics*, Princeton: Princeton University Press.
Carter, B., and Luminet, J.P. 1985, *Mon. Not. R. ast. Soc.*, **212**, 23.
Crane, P., *et al.* 1993, *Astron. J.*, **106**, 1371.
Dressler, A., and Richstone, D.O. 1988, *Astrophys. J.*, **324**, 701.
Duncan, M.J., and Shapiro, S.L. 1983, *Astrophys. J.*, **268**, 565.
Estafthiou, G., *et al* 1988, *Mon. Not. R. ast. Soc.*, **232**, 431.
Frank, J., and Rees, M.J. 1976, *Mon. Not. R. ast. Soc.*, **176**, 633.
Hjellming, M.S., and Webbink, R.F. 1987, *Astrophys. J.*, **318**, 794.
Holdom, B. 1987, *Phys. Rev. D*, **36**, 1000.
Holdom, B. 1993, preprint (UTPT-93-18)hep-ph/9307359.
Holdom, B., and Malaney, R.A. 1994, *Astrophys. J.*, **420**, L53.
Kormendy, J. 1988, *Astrophys. J.*, **325**, 128.
Luminet, J.P., and Pichon, B. 1989, *Astr. Astrophys.*, **209**, 85.
Mao, S., and Paczynski, B. 1992, *Astrophys. J.*, **388**, L45.
Syer, D., Clarke, C.J., and Rees, M.J. 1991, *Mon. Not. R. ast. Soc.*, **250**, 505.
Tremaine, S.D., Ostriker, J.P., and Spitzer, L. 1975, *Astrophys. J.*, **196**, 407.